\documentclass[journal]{IEEEtran}
\usepackage{amsfonts}

\usepackage{cite}
\usepackage{graphicx}
\ifCLASSINFOpdf
\else
\fi
\usepackage[cmex10]{amsmath}
\begin{document}
%
% paper title
% can use linebreaks \\ within to get better formatting as desired
% Do not put math or special symbols in the title.
\title{Controlling Non-reciprocity using enhanced Brillouin Scattering}
%
%
% author names and IEEE memberships
% note positions of commas and nonbreaking spaces ( ~ ) LaTeX will not break
% a structure at a ~ so this keeps an author's name from being broken across
% two lines.
% use \thanks{} to gain access to the first footnote area
% a separate \thanks must be used for each paragraph as LaTeX2e's \thanks
% was not built to handle multiple paragraphs
%

\author{Sameh~Y.~Elnaggar,
         and~Gregory~ N.~Milford% <-this % stops a space
\thanks{S.~Elnaggar and G.~Milford are with the School of Engineering and Information Technology, University of New South Wales, Canberra, ACT, 2600 Australia  e-mails: s.elnaggar@unsw.edu.au and g.milford@adfa.edu.au}% <-this % stops a space
}

\maketitle

% As a general rule, do not put math, special symbols or citations
% in the abstract or keywords.
\begin{abstract}
The properties of space-time modulated media operating in the sub-sonic regime are discussed based on rigorous Bloch-Floquet theory. A geometrical description in the frequency-wavenumber plane is developed to provide insight into the possible interactions and their nature. It is shown that the secular equation has a singularity, which results in a weak/passive second harmonic generation process. Additionally bandgaps arising from the strong/active parametric interaction between an incident wave and its space-time harmonic, result in an inelastic Brillouin like scattering process. Hence when the incident frequency is inside a forward (backward) bandgap, a Stokes' (Anti-Stokes') scattered wave bounces back to the source. Although the forward and backward bandgaps do not generally occur at the same frequency bands, the insertion loss and gap width are equal. Requiring that both gaps do not overlap, enforces a lower bound on the modulation speed. It is shown that although an increase in the modulation index is desirable, as it enhances the non-reciprocal behaviour, it also limits the range of possible modulation speeds. The effective complex refractive index is calculated over a wide frequency range. It is shown that peaks appear in the extinction coefficient, indicating scattering to Stokes' and Anti-Stokes' waves. Finally, a comprehensive numerical analysis based on the Finite Difference Time Domain method is developed to verify and demonstrate the intriguing properties of space-time modulated media.
\end{abstract}

% Note that keywords are not normally used for peerreview papers.
%\begin{IEEEkeywords}
%IEEEtran, journal, \LaTeX, paper, template.
%\end{IEEEkeywords}

% For peer review papers, you can put extra information on the cover
% page as needed:
% \ifCLASSOPTIONpeerreview
% \begin{center} \bfseries EDICS Category: 3-BBND \end{center}
% \fi
%
% For peerreview papers, this IEEEtran command inserts a page break and
% creates the second title. It will be ignored for other modes.
\IEEEpeerreviewmaketitle

\section{Introduction}
% The very first letter is a 2 line initial drop letter followed
% by the rest of the first word in caps.
% 
% form to use if the first word consists of a single letter:
% \IEEEPARstart{A}{demo} file is ....
% 
% form to use if you need the single drop letter followed by
% normal text (unknown if ever used by IEEE):
% \IEEEPARstart{A}{}demo file is ....
% 
% Some journals put the first two words in caps:
% \IEEEPARstart{T}{his demo} file is ....
% 
% Here we have the typical use of a "T" for an initial drop letter
% and "HIS" in caps to complete the first word.
\IEEEPARstart{T}he study of space-time modulated media sparked interest in the late 1950s and early 1960s in the exploration of the properties of travelling wave parametric amplifiers \cite{Tien1958, Cullen1958, Simon1960, Oliner1961, Cassedy1963, Cassedy1967}. In these systems, a strong wave (the pump) modulates a system parameter; for instance a varactor-loaded transmission line can be space-time modulated via the introduction of a strong pump wave that modulates the capacitance value in both space and time. A general solution of the system is determined by a wave and all its infinite space-time harmonics as dictated by the Bloch-Floquet Theorem. Traditionally whenever the modulation index is significantly small, the propagation behaviour can be explained using the \emph{parametric} interaction of  two waves: the signal and idler. This is equivalent to truncating the space-time harmonics expansion to two terms only. This approach, also known as three wave mixing, is widely used to describe nonlinear behaviours in optical media, such as parametric generation and inelastic scattering \cite{Boyd, Fabelinskii}. In an early work, Slater pointed out to the common features between space-time modulated media and scattering from crystals \cite{Slater1958}. Recently we showed that the dynamical behaviour of a sinusoidally pumped nonlinear composite right left handed transmission line (NL CRLH TL) resembles a Stimulated Brilliouin Scattering process observed in crystal structures \cite{Sameh_JAP_NLD, Boyd}.

Despite the success of three wave mixing in describing scattering in nonlinear optical media, its general application to any modulated media may lead to inaccurate results. For example Oliner et al showed that, whenever the speed of modulation is close enough to the speed of the unmodulated medium (in a dispersion-less medium), the full Bloch-Floquet modes must be used \cite{Oliner1961}. They based their arguments on a rigorous mathematical framework that they developed earlier to describe wave propagation in the presence of a spatially modulated surface reactance \cite{Oliner1959}. Very recently we showed that for a NL CRLH TL, three wave mixing does not provide accurate results for strong nonlinearity and/or the TL is relatively short \cite{Sameh_TAP_TWM}.

Quite recently there has been renewed interest in space-time modulated structures for their non-reciprocal behaviour. It has previously been shown that the dispersion relation of a medium loses its symmetry once it is space-time modulated. In this case, waves travelling in the forward and backward directions do not necessary have the same wave number (i.e $\beta^F(\omega)\neq-\beta^B(\omega)$, $F$ and $B$ stand for forward and backward propagation, respectively) \cite{Cassedy1963}. However, such intriguing behaviour has not been exploited until recently. For instance to obtain an optical isolation in one direction, non-reciprocity was introduced via the space-time modulation of the refractive index of a photonic crystal \cite{Fan2009}. This imparts frequency and wave number shifts during a photonic indirect interband transition. The transition is made possible in a given direction by allowing the space-time modulation to phase match the frequencies and wave number of the incident wave and a crystal mode.  This is equivalent to saying that both photons energy and momentum are simultaneously conserved. By properly choosing the length of the crystal to be equal to the coherence length, efficient transfer from the incident wave to its space-time harmonic can be made possible. In the opposite direction of propagation, however, the phase matching conditions are not satisfied; hence a photonic transition does not occur and propagation is not disturbed. The non-reciprocity via inter-band transition can be induced using an electrically driven photonic crystal \cite{Lira2012}.

Using the metamaterial paradigm, space-time modulation was also exploited to introduce non-reciprocity on the meta-atomic level scale by mimicking Faraday's rotation. This was achieved by lifting mode azimuthal degeneracy in a ring resonator via the space time modulation of the dielectric constant \cite{AluCaloz}. The modulation frequency $\omega_m$ allows the clock and anti-clock modes to resonant couple. The coupling process, analyzed by coupled mode theory \cite{Winn1999}, is mediated by the modulated dielectric constant. As a result, the coupled (hybridized) modes are different in character and as a consequence non-reciprocity arises.
 
In essence, the introduction of space-time modulation biases the system, leading to an asymmetric coupling of the space-time harmonics, which results in the non-reciprocal behaviour. This property was recently utilized to break time-reversal symmetry. Such property was exploited to design a multitude of interesting devices; for example: non-reciprocal leaky wave antenna \cite{Hadad2016, Taravati2017mixer}, circulators \cite{Qin2014, Alu2016magnetless}, isolators \cite{Taravati2017oblique} and potential novel devices such as metasurfaces \cite{Caloz2016_STmetasurfaces}.

In the context of an elastic media, it was shown that by properly choosing the modulation speed, a strong interaction of the space-time harmonics can be enabled; this results in the creation of a \emph{directional} band-gap in one direction, while leaving propagation in the reverse direction intact \cite{Trainiti2016}. The directional bandgap identifies a strong active parametric interaction \cite{Oliner1961, Cassedy1963}.

In the current article, we examine the properties of space-time modulated media in terms of the interaction of space-time harmonics from very low frequencies up to frequencies that correspond to the bandgaps, both in the forward and backward directions. The theoretical framework is based on the rigorous continued fraction approach \cite{Oliner1961, Cassedy1963, Cassedy1967}. We present a geometrical description in the $k-\beta$ plane. The detailed analysis delves into the different space-time interactions. We also show that the bandgaps are equivalent to inelastic Brillouin-like scattering centres. The insertion loss due to scattering from a modulated slab is calculated. It is shown that due to the trade-off between the modulation index and the speed of the modulation, a lower bound is enforced on the latter. Brillouin-like scattering and the active space-time harmonics in the bandgaps allow us to explain the abrupt change in the medium extinction coefficient. Finally, a detailed FDTD analysis is developed to verify and demonstrate the intriguing properties of space-time modulated media.

In section II a generalized dispersion relation is derived for an arbitrary space-time periodic medium. For subsequent discussions, the dispersion relation is simplified by considering the fundamental harmonic of the modulation wave only. The analogy to Brillouin scattering is pointed out and the scattering centres are identified. Section III describes in detail the scattering mechanism in both the forward and backward propagation directions, the width of the bandgap as well as the insertion loss in the center of the gap are also estimated; the effect of the inelastic scattering on the extinction coefficient is also demonstrated. In section IV we provide a detailed FDTD analysis to verify the theoretical analysis.
\section{Space-time Dispersion Relation}
Here we will consider the case where the dielectric constant is modulated in both space and time as a travelling wave of the form $\epsilon(z,t)=\epsilon(z-\nu_m t)$, where $\nu_m$ is the modulation speed taken in the $+z$ direction. $\epsilon$ can be expanded in terms of its Fourier harmonics
\begin{equation}
\label{eq:epsilon}
\epsilon=\epsilon_0\sum_{n=-\infty}^{\infty} \epsilon_n'\cos(n\omega_mt-n\beta_mz),
\end{equation}
where $\omega_m$ and $\beta_m$ are the temporal and spatial modulation frequencies, and are related to the modulation speed $\nu_m$ as:
\begin{equation}
\nu_m=\frac{\omega_m}{\beta_m}.
\end{equation}

\begin{figure}
\centering
\includegraphics[width=2.0in]{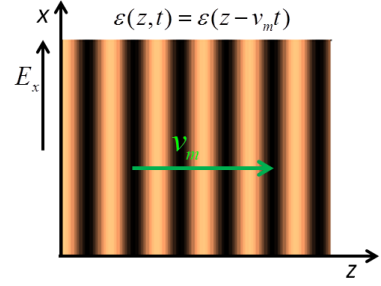}
\caption{A  space-time modulated medium.}
\label{fig:OneDProblem}
\end{figure}

Seeking an $x$ polarised TEM solution for the wave propagating along the $z$ axis (Fig. \ref{fig:OneDProblem}), the wave equation can be written as
\begin{equation}
\label{eq:DE}
\frac{\partial^2 E_x(z,t)}{\partial z^2}=\mu_0\frac{\partial^2 \epsilon(z,t)E_x(z,t)}{\partial t^2}.
\end{equation}
The disturbance $\epsilon(z,t)$ in (\ref{eq:DE}) is periodic in both time and space, hence we can apply Bloch-Floquet theorem to express the solution as a propagating wave $E_0\exp(-j[\omega t-\beta z])$ modulated by a space-time periodic function $P(\omega_m t-\beta_m z)$; where $\omega$ and $\beta$ are the yet to be determined frequency and wave number. The periodic function $P$ can  in turn be written as an infinite Fourier series:
\begin{equation}
P(\omega_m t-\beta_m z)=\sum_{r=-\infty}^\infty a_r e^{-jr(\omega_m t-\beta_m z)}.
\end{equation}
Therefore, the general solution assumes the form:
\begin{equation}
\label{eq:gen_sol}
E_x(z,t)=\sum_{r=-\infty}^\infty a_r e^{-j[(\omega+r\omega_m) t-(\beta+r\beta_m) z]},
\end{equation}
where the arbitrary constant $E_0$ is absorbed in $a_r$.
Substituting (\ref{eq:gen_sol}) and (\ref{eq:epsilon}) in (\ref{eq:DE}); and noting that $\cos\theta=(\exp(j\theta)+\exp(-j\theta))/2$, a recursion relation between the $a_r$ terms can be found to be 
\begin{equation}
\label{eq:recursionGeneral}
\sum_{s=1}^{\infty}\epsilon_s\left(a_{r-s}+a_{r+s}\right)+F_r(k,\beta)a_r=0,~~r\in \mathbb{Z}
\end{equation}
where 
\begin{equation}
\label{eq:F}
F_r(k,\beta)=2\left(1-\left[\frac{\beta a+2\pi r}{ka+2\pi\nu r}\right]^2\right),
\end{equation}
$a=2\pi/\beta_m$ the wavelength of the modulation, $k$ is the unmodulated wave number ($k=\omega/c$), $c$ is the speed of the unmodulated medium; and $\nu=\nu_m/c$ is the relative speed of the modulation. (\ref{eq:recursionGeneral}) represents an infinite system of linear algebraic equations that couple the $r^\textnormal{th}$ space-time (Floquet) mode to all other modes. To find the dispersion relation for a given $k$ ($\beta$), a secular (or characteristic) equation is obtained by setting the determinant of the infinitely countable system (\ref{eq:recursionGeneral}) to zero; hence the corresponding $\beta$ ($k$) can be calculated. When $v_m=0$, $F_{-r}(k,-\beta)=F_r(k,\beta)$. Therefore the system of equations (\ref{eq:recursionGeneral}) is invariant under the reflection of $r$ ($r\Rightarrow -r$), which means that if ($k$, $\beta$) is a solution to the characteristic equation so does ($k$, $-\beta$) and the modulated medium is reciprocal. However for a general $\nu_m\neq 0$, this is not the case and the medium is intrinsically non-reciprocal.

In practical situations, the expansion (\ref{eq:epsilon}) can be truncated to the fundamental component ($\omega_m$ and $\beta_m$) only, which simplifies (\ref{eq:recursionGeneral}) to
\begin{equation}
\label{eq:recursion}
a_{r+1}+D_ra_r+a_{r-1}=0,
\end{equation}
for $r \in \mathbb{Z}$ and
\begin{equation}
\label{eq:Dn}
D_r=\frac{F_r(k,\beta)}{M}=\frac{2}{M}\left[1-\left(\frac{\beta a+2\pi r}{ka+2\pi\nu r}\right)^2\right],
\end{equation}
where  $M=\epsilon_1'/\epsilon_0'$ is the modulation index \cite{Oliner1961}. The modulated media then couples the $r^\textnormal{th}$ Floquet mode with its nearest neighbour (+1 and -1 harmonics) only.

Formally, a rigorous \emph{continued fractions} approach was derived to determine the dispersion relation \cite{Oliner1961, Cassedy1963}. Expressing the ratio $a_r/a_{r-1}$ as a continued fraction and noting that $a_r/a_{r-1}=(a_{r-1}/a_r)^{-1}$, the secular equation can be cast into the form of a continued fraction form \cite{Oliner1961, Cassedy1963}
\begin{multline}
\label{eq:characteristic}
G_r(ka,\beta a)\equiv
\\ D_r-\frac{1}{D_{r-1}-\frac{1}{D_{r-2}-\frac{1}{\ddots}}}-\frac{1}{D_{r+1}-\frac{1}{D_{r+2}-\frac{1}{\ddots}}}=0.
\end{multline}

It is worth noting that $G_r(ka,\beta a)=G_0(ka+2\pi\nu r, \beta a+2\pi r)$, which means that the dispersion relation can be fully obtained from any of the infinite $G_r(ka,\beta a)$ and they are all compatible.

For infinitesimally small $M\rightarrow 0$, the equations  (\ref{eq:recursion}) decouple and the dispersion relation reduces to
\begin{equation}
\label{eq:MsmallDispersion}
MD_r=0\Rightarrow \beta+r\beta_m=\pm\left(\omega+r\omega_m\right)/c,
\end{equation}
the dispersion relation of the unmodulated medium, but shifted by $(r\beta_m, r\omega_m)$.

For each frequency $\omega$, (\ref{eq:characteristic}) determines the corresponding wave number $\beta$. Therefore, the dispersion relation can be constructed, which is generally a function of $M$ and $\nu$. However to guarantee that $a_r/a_{r-1}$ converges for some $r>r_0$, it was shown that $|D_r|$ must be greater than 2; this is equivalent to saying that the modulation velocity $v_m$ may not be very close to the speed  $c$ of waves in the unmodulated medium \cite{Oliner1961, Cassedy1963}. Moreover, for stable operation (all solutions bounded in time), the sub-sonic condition $v_m<c$ or equivalently $\nu<1$ must be met. This enforces an upper bound on the modulation speed \cite{Cassedy1967}.

\subsection{Brillouin Scattering Analogy}
Equation (\ref{eq:DE}) establishes the formal connection between space-time modulation and Brillouin Scattering. However the analogy between the two needs more discussion. Brillouin Scattering stems from the inevitable fluctuation of thermodynamic variables, which affects the macroscopic properties of the crystal lattice. However such fluctuation is random, reciprocal and wide band \cite{Fabelinskii}. Nevertheless, interaction is usually negligible except for a specific crystal mode: the one that satisfies the phase matching conditions \cite{Fabelinskii, Boyd}. Additionally, the interaction is weak, with a very small relative speed $\nu\sim 1~ \textnormal{ppm}$. Space-time modulated media can then be regarded as a \emph{Stimulated} or \emph{Engineered} Enhanced Brillouin Scattering. Such an analogue behaviour was recently exploited to describe the interaction of a longitudinal acoustic wave with a spatio-temporal phononic crystal \cite{Croenne2017brillouin}.
\subsection{Scattering Centres}
\begin{figure}
\centering
\includegraphics[width=2.75in]{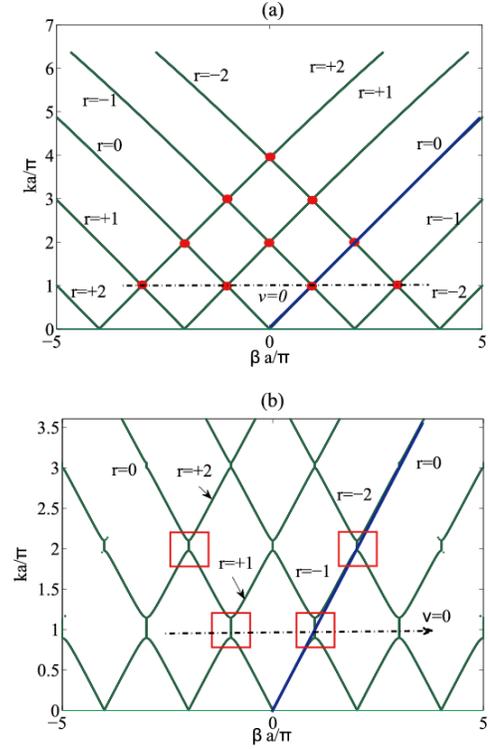}
\caption{Dispersion Characteristics for a space only modulated medium ($\nu=0$) (a) negligibly small values of $M$. The forward fundamental branch $\omega=\beta c$ is highlighted. The dots represent the scattering centres where interactions are possible. Scattering centres in the forward (backward) direction correspond to Stokes (Anti-Stokes) centres. (b)  $M=0.5$; the forward fundamental branch $\omega=\beta c$ is highlighted. The rectangles identify the regions of strong harmonic scattering.}
\label{fig:Dispersion_vm0}
\end{figure}

For small values of $M$, the dispersion relation approaches (\ref{eq:MsmallDispersion}). Figure \ref{fig:Dispersion_vm0}(a) depicts the dispersion relation for a space only modulation medium ($\nu=0$) and an infinitesimally small value of $M$. Comparing this to Fig. \ref{fig:Dispersion_vm0}(b), which depicts the dispersion relation but for $M=0.5$, (\ref{eq:characteristic}), reveals that the intersection points between the $MD_r=0$ lines identify the loci of strong interactions. As will be shown, for a general $\nu\neq 0$ these points allow the incident wave to scatter in an inelastic fashion and therefore will be called the \emph{Scattering Centres}. The behaviour of such interactions can be studied by considering the interaction with the $r=0$ branches only. In the forward direction ($\beta>0$), the intersection point of the $r=0$ branch and another arbitrary $r\neq 0$ branch can be found to be
\begin{equation}
\label{eq:ffreq}
k_{r}^Fa=\beta_{r}^Fa=r\pi\left(1+\nu\right)~~\textnormal{for}~r<0,
\end{equation}
where the subscript $r$ identifies that this is the intersection with the $r^\textnormal{th}$ branch and the superscript $F$ emphasises that this is for a forward propagating wave (along $+z$ axis). Similarly $k_{r}^B$ and $\beta_{r}^B$, the intersection with the backward branch is found to be
\begin{equation}
\label{bfreq}
k_{r}^Ba=-\beta_{r}^Ba=r\pi\left(1-\nu\right)~~\textnormal{for}~r>0.
\end{equation}

\begin{figure}
\centering
\includegraphics[width=2.75in]{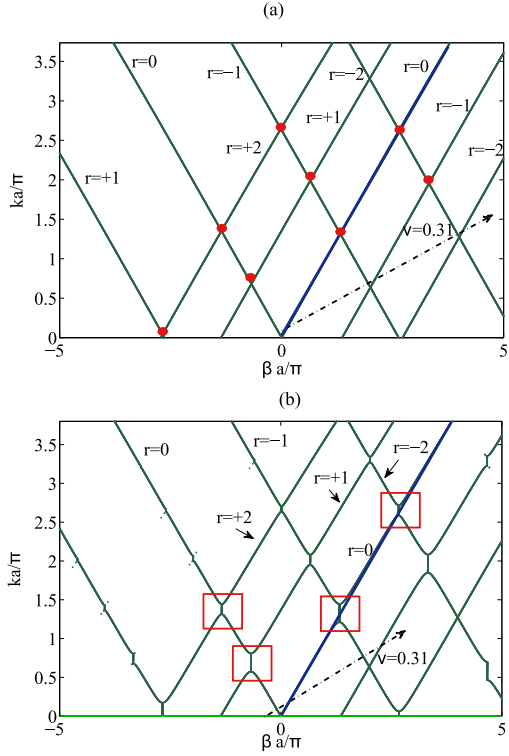}
\caption{Dispersion Characteristics for a space-time modulated medium ($\nu=0.31$) (a) negligibly small values of $M$. The forward fundamental branch $\omega=\beta c$ is highlighted. The dots represent the scattering centres where interactions are possible. Scattering centres in the forward (backward) direction correspond to Stokes (Anti-Stokes) centres. (b)  $M=0.5$; the forward fundamental branch $\omega=\beta c$ is highlighted. The rectangles identify the regions of strong harmonic scattering.}
\label{fig:Dispersion_vm0p31}
\end{figure}

For reasons that will be revealed in Section III, we call the forward and backward scattering centres Stokes' and Anti Stokes' centres, respectively. For a forward propagating modulation, $\nu\geq 0$ and hence $k_{r}^B\leq k_{r}^F$; the equality holds for $\nu=0$: space only modulated medium. The inequality $k_{r}^B< k_{r}^F$ means that the interaction in the backward branch occurs at a lower frequency leading the medium to become non-reciprocal. Fig. \ref{fig:Dispersion_vm0p31} presents the dispersion relation for $\nu=0.31$ and two values of $M$:  $M=0$ and $M=0.5$. It is clear that $k_{r}^B< k_{r}^F$.

\section{Scattering Mechanism}
\begin{figure}
\centering
\includegraphics[width=2.8in]{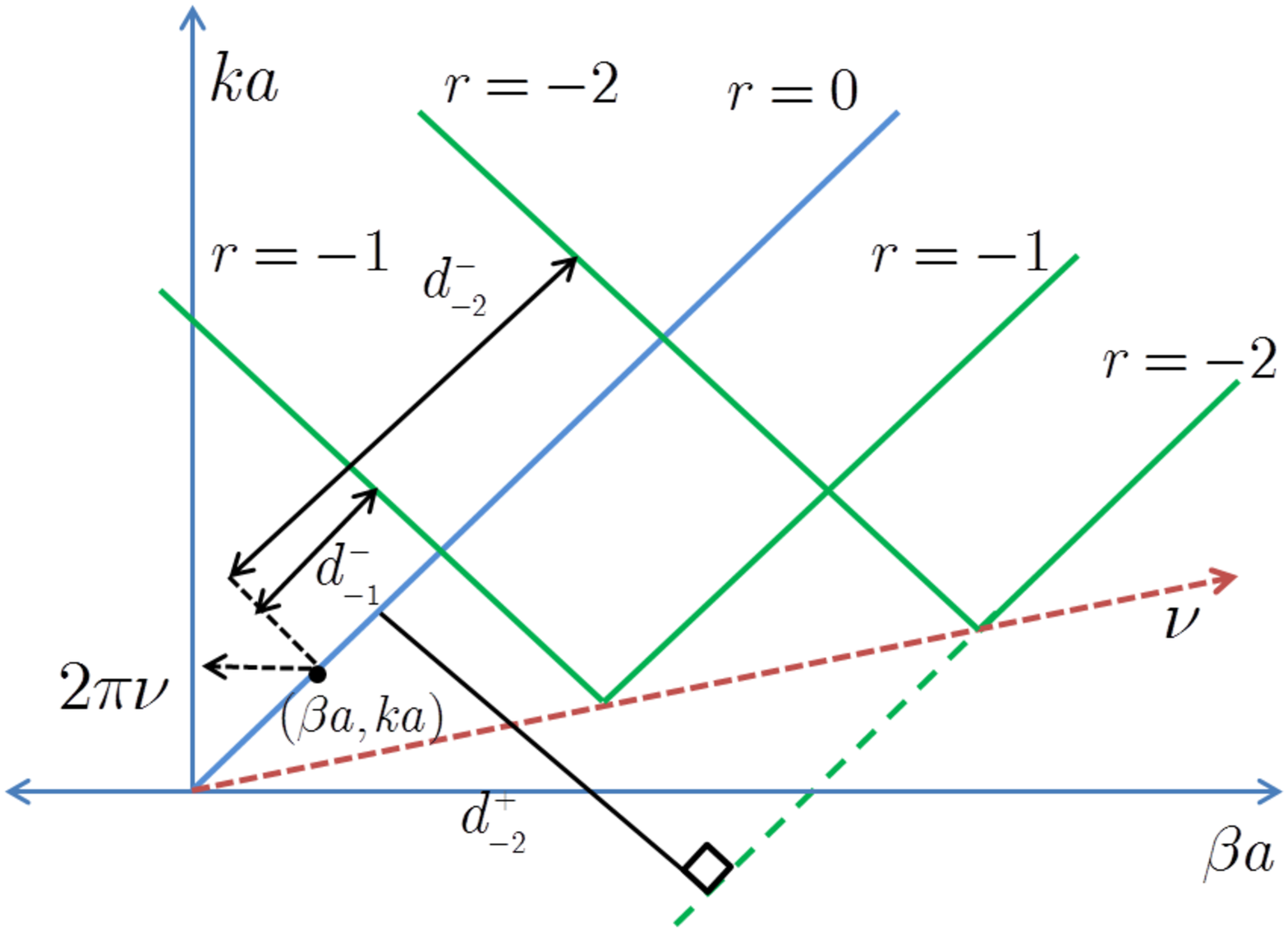}
\caption{Geometrical Description of the dispersion characteristics.}
\label{fig:DrGeometry}
\end{figure}
In this section we describe the conversion (scattering) process from the fundamental wave with frequency $\omega$ and wave number $\beta$ to its space-time harmonics with frequency $\omega+r\omega_m$ and wave number $\beta+r\beta_m$. The scattering behaviour depends on the values of $D_r$. First, we examine the trend of $D_r$ as a function of the input frequency $\omega$ or the corresponding scaled value $ka$. Toward this end, $D_r$ is related to metric distances on the $ka-\beta a$ plane. Referring to Fig. \ref{fig:DrGeometry}, the distances $d_{r}^+$, between an arbitrary point $(\beta a, ka)$ on the $r=0$ line and the branch of $D_r=0,~r\neq 0$ with a positive slope is 
\begin{equation}
d_{r}^+=\frac{|(\beta a+2\pi r)-(ka+2\pi\nu r)|}{\sqrt{2}}=\frac{2\pi|r|(1-\nu)}{\sqrt{2}},
\end{equation}
which is obtained by substituting the coordinate $(\beta a=ka, ka )$ in the expression $ka+2\pi\nu r-\beta a-2\pi r$ and dividing by $\sqrt{2}$. Similarly, the distance $d_{r}^-$ between $(\beta a, ka)$ and the branch of $D_r=0$ with a negative slope ($ka+2\pi\nu r=-(\beta a+2\pi r)$) is
\begin{equation}
d_{r}^-=\frac{|(\beta a+2\pi r)+(ka+2\pi\nu r)|}{\sqrt{2}}.
\end{equation}
Therefore,
\begin{equation}
\label{eq:Dr_geometry}
M|D_r|=\frac{4d_{r}^+d_{r}^-}{(ka+2\pi r\nu)^2}.
\end{equation}
The above Eq., together with Fig. \ref{fig:DrGeometry},  give a pictorial view of how $|D_r|$ depends on the geometrical metrics $d_{r}^+$ and $d_{r}^-$. For very small values of $\beta a$ and $ka$, $|D_r|=2/M(1-\nu^2)/\nu^2$ and is independent of $r$; this means that for very low frequencies all $|D_r|$ basically have the same value. Nevertheless as $ka$ increases, $D_r$ effect is radically different for waves travelling in the forward or backward directions, as will be detailed in the next two subsections.

\subsection{Forward direction ($\beta>0$)} 
\begin{figure}
\centering
\includegraphics[width=3.0in]{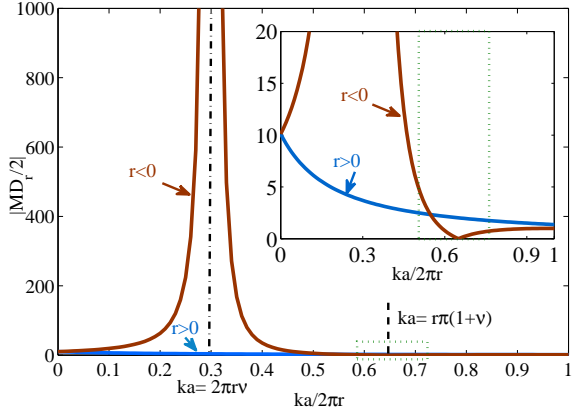}
\caption{Normalized $D_r$ as a function of the normalized frequency $ka$ for the forward propagation.}
\label{fig:Drforward}
\end{figure}
For a wave travelling in the $+z$ direction and assuming that propagation is mainly due to the $r=0$ branch, $\beta a=ka$, which is true for small values of $M$ and away from the bandgaps highlighted in Figs. \ref{fig:Dispersion_vm0}(b) and \ref{fig:Dispersion_vm0p31}(b). Nevertheless, the general behaviour of $D_r$ around the bandgaps can still be described using the $\beta a=ka$ approximation; for accurate results we resort to the general secular equation (\ref{eq:characteristic}). Fig. \ref{fig:Drforward} shows the change of $D_r$ as a function of the normalized frequency $ka/2\pi r$. For $r<0$ ($r>0$), as $ka$ increases along the $r=0$ line, $d_{r}^-$ linearly decreases (increases) and $d_{r}^+$ remains constant (refer to Fig. \ref{fig:DrGeometry}). Additionally, the denominator of (\ref{eq:Dr_geometry}) quadratically decreases (increases). Therefore, for $r>0$,  the net effect of both terms keeps $|D_r|$ monotonically decreasing.

The trend of $|D_r|$ for $ r<0$ is more dynamic; as $ka$ approaches the singularity at $-2\pi r\nu$, $|D_r|$ increases  without bound. However, as $ka$ increases further, $|D_r|$ will decrease and eventually vanish at the Stokes' centre $ka=r\pi(1+\nu)$. At this point the $r^\textnormal{th}$ branch intersects the main branch (i.e, $d_{r}^-=0$) and the interaction is the strongest, as was already shown in Figs. \ref{fig:Dispersion_vm0} and \ref{fig:Dispersion_vm0p31}. For such cases where $|D_r|$ attains small values, the continued fraction in (\ref{eq:characteristic}) has to be used to determine $\beta a$. Otherwise, the dispersion characteristic is determined via $D_0=0$ alone, which is equivalent to the approximation used here ($ka=\beta a$).

At the first singular point $ka=2\pi\nu $, the secular equation (\ref{eq:characteristic}) reduces to
\begin{equation}
D_0-\frac{1}{D_{+1}-\frac{1}{D_{+2}-\frac{1}{\ddots}}}=0,
\end{equation}
i.e, it depends on the interaction between the main branch and the positive $r$ space time harmonics. To find the properties and strength of this interaction, it is assumed that at $ka=2\pi\nu$, $D_{r} $ for $r>1$ are large enough such that their contribution to the secular equation can be ignored. This is consistent with the monotonically decreasing trend of $|D_r|,~r>0$ (Fig. \ref{fig:Drforward}). One can then solve the truncated equation:
\begin{equation}
\label{eq:SHGcharacteristic}
D_0D_{+1}-1=0
\end{equation}
to find $\beta a$, evaluated at $ka=2\pi\nu$. Since $\beta a$ is not very different from its value at $r=0$, it can be approximated by
\begin{equation}
\beta a=2\pi\nu+\eta,~~~|\eta|\ll 1.
\end{equation}
Neglecting orders in $\eta$ higher than two, the truncated characteristic equation (\ref{eq:SHGcharacteristic}) is reduced to 
\begin{equation}
\eta^2\left[\frac{\gamma^2-1}{\left(2\pi\nu\right)^2}+\frac{2\gamma}{\left(2\pi\nu\right)^2}\right]+\eta\frac{\gamma^2-1}{\pi\nu}-\left(\frac{M}{2}\right)^2=0,
\end{equation}
where $\gamma=(\nu+1)/2\nu$ and is always greater than one (since $\nu<1$). The type of the solution is determined by the radical, which for the case in hand, is always positive. Hence, $\eta$ is real, indicating a passive exchange of power between the wave at normalized frequency $ka=2\pi\nu$ and its space-time harmonic $ka+2\pi\nu=2ka$ \cite{Louisell1960}. This interaction corresponds to a weak second harmonic generation. To estimate the magnitude of the interaction (i.e, finding $\eta$), one notices that for $\eta^2\ll |\eta|$ 
\begin{equation}
\eta\approx \frac{\pi\nu^3M^2}{1+2\nu-3\nu^2},
\end{equation}
a second order in $M$. Therefore
\begin{equation}
\label{eq:eta_forward_passive}
\beta a\approx2\pi\nu\left(1+\frac{M^2\nu^2}{2\left(1+2\nu-3\nu^2\right)}\right).
\end{equation}
The ratio of the amplitude of the +1 harmonic and the fundamental can be determined as
\begin{equation}
\label{eq:a1_a0_SHG}
\frac{a_1}{a_0}=-D_0\approx\frac{2M\nu^2}{1+2\nu-3\nu^2}.
\end{equation}
Substituting with the typical values, $M=0.5$, $\nu=0.3$, $a_1/a_0\approx 0.07$, a very small number. Therefore, it can be concluded the the interaction between the fundamental and its +1 harmonic is very weak and can be neglected in the cases of interest. This conclusion is consistent with the dispersion relations plotted in Figs. \ref{fig:Dispersion_vm0}(b) and \ref{fig:Dispersion_vm0p31}(b), where at $ka=2\pi\nu$, the dispersion relation is basically that of the unmodulated medium.

On the other hand, the interaction at $ka=r\pi(1+\nu)$ between the fundamental and its $-|r|$ harmonic is much stronger. In general the first interaction (with the -1 harmonic) is the strongest and provides the wider bandwidth. For this reason and because it is straight forward to find closed form expressions, we will restrict our attention to the -1 harmonic interaction. At $ka=r\pi(1+\nu)$, $D_{-1}$ is small (inset of Fig. \ref{fig:Drforward}) to the extent that the dispersion relation (\ref{eq:characteristic}) can be truncated to
\begin{equation}
\label{eq:SecularForwardSimplified}
D_0D_{-1}-1=0
\end{equation}
Letting $\beta a=ka+\eta=\pi(1+\nu)+\eta$, where $|\eta|\ll\pi(1+\nu)$, $\eta$ can be approximated to
\begin{equation}
\label{eq:Eta_Fwd}
\eta=\pm j\pi(1-\nu^2)^{1/2}M/4.
\end{equation}
In terms of the wave number
\begin{equation}
\beta=k\left(1\pm j\alpha_F\right),
\end{equation}
where 
\begin{equation}
\alpha_F=\frac{M}{4}\sqrt{1-\frac{\omega_m}{\omega}}=\frac{M}{4}\sqrt{\frac{1-\nu}{1+\nu}}.
\end{equation}
Unlike (\ref{eq:eta_forward_passive}), $\eta$ is imaginary and first order in $M$, indicating a strong active interaction \cite{Louisell1960}. The amplitude of the -1 harmonic can be found to be
\begin{equation}
\frac{a_{-1}}{a_0}=-D_0=\pm j\sqrt{\frac{1-\nu}{1+\nu}}\equiv \pm jb,
\end{equation}
which, unlike (\ref{eq:a1_a0_SHG}), does not depend on $M$.

The general solution can be written as follows \cite{Simon1960}
\begin{equation}
E_x=a_+E_{x+}+a_-E_{x-},
\end{equation}
where
\begin{equation}
\label{eq:Ex}
E_{x\pm}=\left(e^{-j\left(\omega t-kz\right)}\pm jbe^{-j\left(\left[\omega-\omega_m\right]t-\left[k-\beta_m\right]z\right)}\right)e^{\pm\alpha_Fkz},
\end{equation}
Please note that the above general solution assumes that the contributions from other branches are ignored (ignoring the multi-valued character of $k(\omega)$). Noting that \cite{Simon1960} 
\begin{equation}
\label{eq:phasesynchronismforward}
\frac{\omega}{k}=-\frac{\omega-\omega_m}{k-\beta_m},
\end{equation}
 the characteristic impedances of the space-time medium at frequencies $\omega$ and $\omega-\omega_m$ are
\begin{equation}
\label{eq:Z0_forward}
Z_0=\frac{\bar{Z}}{1\mp j\alpha_F }
\end{equation}
and 
\begin{equation}
\label{eq:Z1_forward}
Z_1=\frac{-\bar{Z}}{1\pm j\alpha_F/\left(1-\omega_m/\omega\right)},
\end{equation}
respectively, and $\bar{Z}=\omega\mu_0/k$ is the impedance of the unmodulated medium.

According to (\ref{eq:phasesynchronismforward}), the fundamental and scattered waves have phase velocities which are equal in magnitude but opposite in direction. Moreover, the sign of $Z_1$ is negative implying that the scattered wave is travelling in the $-z$ direction. This is consistent with Coupled Mode Theory predictions where active interaction is possible between waves which have the same magnitude of phase velocity and energy flow is contra-directive \cite{Pierce1954}. Additionally, the phase matching conditions automatically emerge:
\begin{eqnarray}
\label{eq:PM_forward}
\omega_S=\omega-\omega_m\\
\beta_S=\beta-\beta_m.
\end{eqnarray}
Since the scattered wave is red-shifted (lower in frequency), it corresponds to a Stokes' wave, which explains why the scattering centre was named Stokes' centre earlier in Section II.

\begin{figure}
\centering
\includegraphics[width=2.7in]{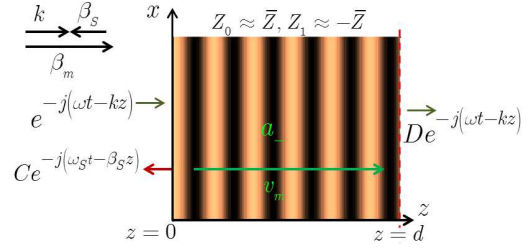}
\caption{Simplified boundary value problem for an incident wave that is co-directional with the modulation wave.}
\label{fig:ForwardBVP_simplified}
\end{figure}

Considering the situation depicted in Fig. \ref{fig:ForwardBVP_simplified}, where a wave of frequency $\omega$ impinges the modulated medium of length $d$.  The tangential fields $E_x$ and $H_y$ are continuous at the interfaces $z=0$ and $z=d$.  Approximating the impedance of the fundamental and +1 harmonic to $\bar{Z}$ and $-\bar{Z}$, respectively as shown in the Fig., the amplitude $D$ of the transmitted wave is found to be
\begin{equation}
D=e^{-\alpha_Fk_0d}
\end{equation}
and the total attenuation, $\textnormal{ATT}$, of the modulated medium in Nepers is
\begin{equation}
\label{eq:ATT_forward}
\textnormal{ATT}=\alpha_Fkd=\frac{\pi}{4}M\sqrt{1-\nu^2} \frac{d}{a},
\end{equation}
which reveals interesting conclusions. As expected the attenuation is directly proportional to $M$, but decreases as the modulation speed $\nu_m$ increases. It is then desirable to make $\nu_m$ as small as possible. However, as will be shown in the subsection C, $\nu_m$ has a lower bound determined by non-reciprocity. Additionally, from (\ref{eq:ATT_forward})  the attenuation is proportional to the normalized length of the medium (normalized to the modulation wavelength); hence suggesting that to have an efficient scattering the modulation wavelength $a$ should be as small as possible. However, this might be constrained by the lower or upper bound of $\nu_m$.

\subsection{Backward Direction ($\beta<0$)}
For the backward propagation $\beta<0$, the $D_r$ values along the $\beta a=-ka$ line assume the form
\begin{equation}
\label{eq:normalizedDr_BWD}
\left| \frac{MD_r}{2}\right|=\left|1-\left(\frac{1-ka/2\pi r}{\nu+ka/2\pi r}\right)^2\right|.
\end{equation}

\begin{figure}
\centering
\includegraphics[width=3.0in]{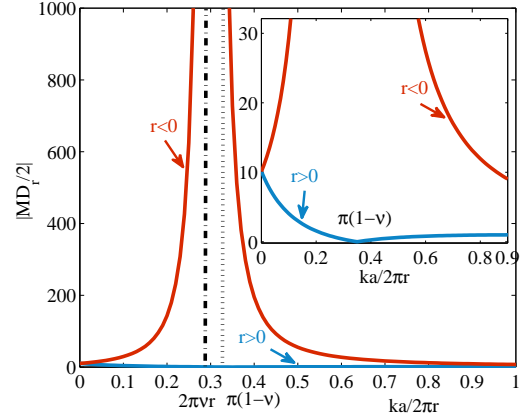}
\caption{Normalized $D_r$ as a function of the normalized frequency $ka$ for the backward propagation.}
\label{fig:Drbackward}
\end{figure}

Fig. \ref{fig:Drbackward} shows how $D_r$ changes as a function of the input frequency.  Although the $D_r$ trend in Fig. \ref{fig:Drbackward} look similar to Fig. \ref{fig:Drforward}, there are some fundamental differences which do not allow the backward propagation to be treated as the mere dual of the forward one. First, for the forward propagation, at the singularity $ka=2\pi\nu$ the interaction is with the  $r>0$ harmonics only. As a result, it was shown in the previous subsection that the +1 harmonic interacts passively with the fundamental. Additionally, it is not possible to position the singularity at the Stokes' scattering centre $ka=\pi(1+\nu)$. However for backward propagation, the singularity is still at $ka=2\pi\nu$. Additionally, choosing $\nu=1/3$ positions the singularity at the Anti-Stokes' centre, hence limiting the scattering to be strictly with the $r>0$ harmonics only. 
Similar to the analysis of the previous subsection, the truncated secular equation
\begin{equation}
\label{eq:SecularBackwardSimplified}
D_0D_{-1}-1=0
\end{equation}
can be solved to find the propagation constant $\beta a=-ka+\eta$ at $ka=\pi(1-\nu)$. Neglecting terms of order higher than $\eta^2$, $\eta$ is found to be
\begin{equation}
\eta=\pm j\pi\left(1-\nu^2\right)^{1/2}\frac{M}{4},
\end{equation}
identical to (\ref{eq:Eta_Fwd}). In terms of the wave numbers
\begin{equation}
\label{eq:k_Bwd}
k=k_0\left(1\pm j\alpha_B\right),
\end{equation}
where 
\begin{equation}
\label{eq:alphaB}
\alpha_B=\frac{M}{4}\sqrt{1+\frac{\omega_m}{\omega}}.
\end{equation}
Similarly,
\begin{equation}
\label{eq:a1_a0_Bwd}
\frac{a_{1}}{a_0}=\pm j\sqrt{\frac{1+\nu}{1-\nu}}\equiv \pm jb
\end{equation}

\begin{figure}
\centering
\includegraphics[width=2.7in]{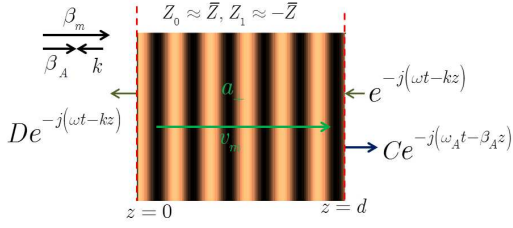}
\caption{Simplified boundary value problem for an incident wave travelling in a direction opposite to the modulation.}
\label{fig:BackwardBVP_simplified}
\end{figure}

For small values of $M$, Fig. \ref{fig:BackwardBVP_simplified} shows the situation where the modulated medium is impinged by a wave at frequency $\omega$. It is worth to notice that the two situations depicted in Figs. \ref{fig:ForwardBVP_simplified} and \ref{fig:BackwardBVP_simplified} represent waves which have different frequencies. In the forward direction the wave is at a normalized frequency $\pi(1+\nu)$, while in the backward direction its normalized frequency is $\pi (1-\nu)$. 

The backward attenuation takes a form identical to (\ref{eq:ATT_forward}), but at a normalized frequency of $ka=\pi(1-\nu)$. This means that by the proper selection of $M$ and $\nu$ the medium acts as non-reciprocal bandstop filter, where the forward and backward stop bands can occur at different frequency ranges.

The scattered wave is an Anti-Stokes wave and the interaction automatically satisfies the phase matching conditions
\begin{eqnarray}
\label{eq:PMbackward}
\omega_A=\omega+\omega_m\\
\beta_A=\beta+\beta_m.
\end{eqnarray}

\subsection{Width of Directional Bandgap}
\begin{figure}
\centering
\includegraphics[width=3.0in]{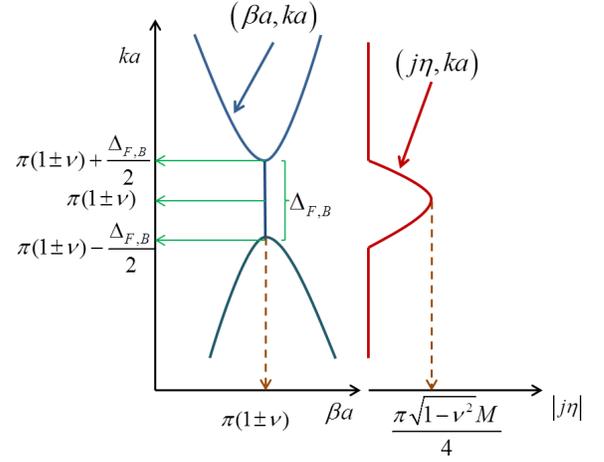}
\caption{Normalized frequency versus propagation and attenuation constants at the vicinity of the bandgap.}
\label{fig:Bandgap}
\end{figure}

The width of the directional bandgap determines the bandwidth at which the medium exhibits strong non-reciprocity. It also sets the lower bound on the modulation speed $\nu$. To demonstrate non-reciprocity in one direction only, the forward and backward bandgaps must not overlap. Hence
\begin{equation}
\label{eq:InequalityBandGap}
\frac{\Delta_F+\Delta_B}{2}<2\pi\nu,
\end{equation}
where $\Delta_F$ and $\Delta_B$ are the directional bandgaps in the forward and backward directions, respectively. At the band edges $\beta a=\pi(1\pm\nu)$ (Fig. \ref{fig:Bandgap}). Substituting these values in the corresponding secular equations (\ref{eq:SecularForwardSimplified}) and (\ref{eq:SecularBackwardSimplified}), it can be found that to second order:
\begin{equation}
\Delta_F=\Delta_B=\pi\left(1-\nu^2\right)^{1/2}\frac{M}{2}.
\end{equation}
Therefore, the minimum value of $\nu$ is determined from the inequality (\ref{eq:InequalityBandGap}) to be
\begin{equation}
\nu_{\textnormal{min}}=M/4,
\end{equation}
which is identical to the value determined in \cite{Trainiti2016} for modulated elastic media. Therefore for optimum non-reciprocal behaviour
\begin{equation}
\frac{M}{4}<\nu<1.
\end{equation}
 
\subsection{Complex Refractive Index, $\tilde{n}=n+j\kappa$}
\begin{figure}
\centering
\includegraphics[width=3.5in]{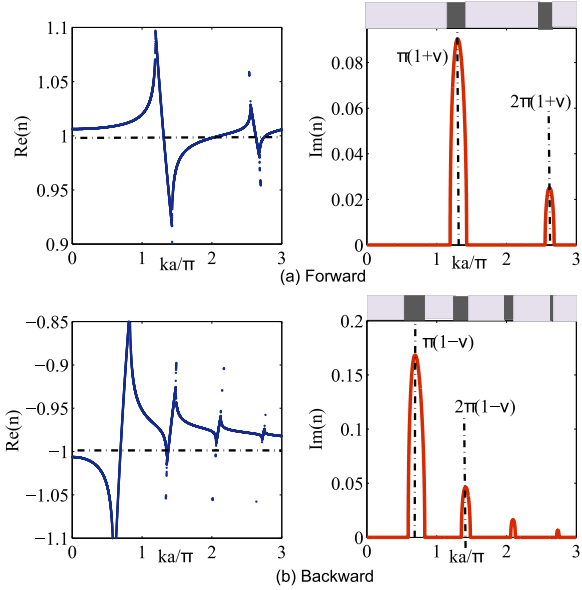}
\caption{Refractive index and extinction coefficient of waves having normalized frequency $ka$. (a: Forward direction. (b): Backward direction (the real part of the refraction coefficient is negative to emphasize that the wave is travelling in the $-z$ direction). The dark and bright bands on top of the extinction coefficient plots depict the situations when the medium is opaque or transparent, respectively.}
\label{fig:RefractiveIndex}
\end{figure}

At an arbitrary incident frequency $\omega$, the complex wavevector $\tilde{k}$ can be written as
\begin{equation}
\tilde{k}=\tilde{n}k,
\end{equation}
where $\tilde{n}=n+j\kappa$ is the complex refractive index. The imaginary part determines the efficiency of the power scattered by the space-time harmonics (forward: $\omega-\omega_m$, backward $\omega+\omega_m$). Using the dispersion relation (\ref{eq:characteristic}), the refractive index is calculated for the forward and backward directions as shown in Fig. \ref{fig:RefractiveIndex}. For the calculations, 20 terms were used to compute the continued fractions; usually four of five terms are enough, as the continued fractions rapidly converge.

From Fig. \ref{fig:RefractiveIndex}, it is clear that the optical property of the medium is non-reciprocal; absorption occurs at different input frequencies. Additionally, the widths of the Stokes and Anti Stokes Centres are basically the same because $\Delta_F= \Delta_B$ as was already determined in the previous subsection.

\section{FDTD Analysis}
\begin{figure}
\centering
\includegraphics[width=3.5in]{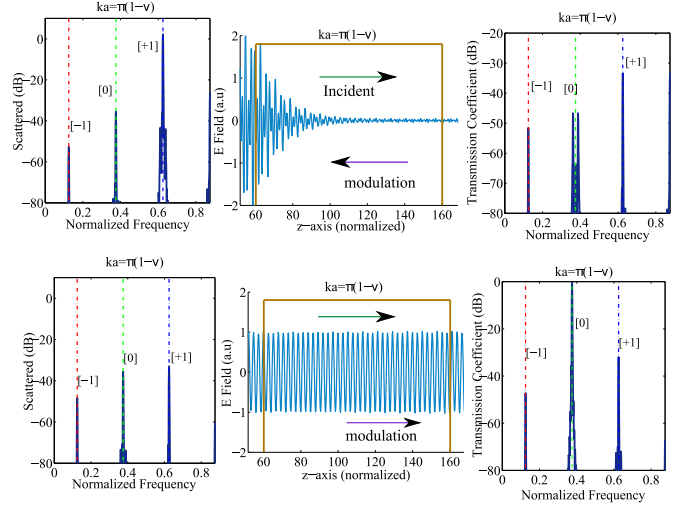}
\caption{FDTD simulated spectra for the scattered and transmitted waves for $\pi(1-\nu)$. The modulation speed changed direction and the system exhibits a non-reciprocal behaviour. ([0]: incident wave at $\omega$, [-1]: $\omega-\omega_m$, [+1]: $\omega+\omega_m$).}
\label{fig:Nonrecip_bwd}
\end{figure}

The space-time dependence of the permittivity was used to modify the update equations of the FDTD formalism to determine the propagation and scattering behaviour of a space-time modulated medium. For more details on the FDTD implementation please refer to the Appendix. Fig. \ref{fig:Nonrecip_bwd}(a) presents the scenario where $ka=\pi(1-\nu)$ and the modulation travels in the $-z$ direction, In this case, the incident wave interacts with its +1 harmonic, which scatters energy back in the $-z$ direction, as is clear after inspecting the frequency spectrum of the scattered field. At this point, the scattered wave is the Anti-Stokes' wave in Brillouin Scattering. However if the modulation speed was inverted as in Fig. \ref{fig:Nonrecip_bwd}(b), the modulation and incident waves are co-directional. In this case, as shown in Fig. \ref{fig:Nonrecip_bwd}(b), propagation is not disturbed and the medium is transparent. Hence, the medium is non-reciprocal at $ka=\pi(1-\nu)$, in agreement with the analytical predictions.

\begin{figure}
\centering
\includegraphics[width=3.5in]{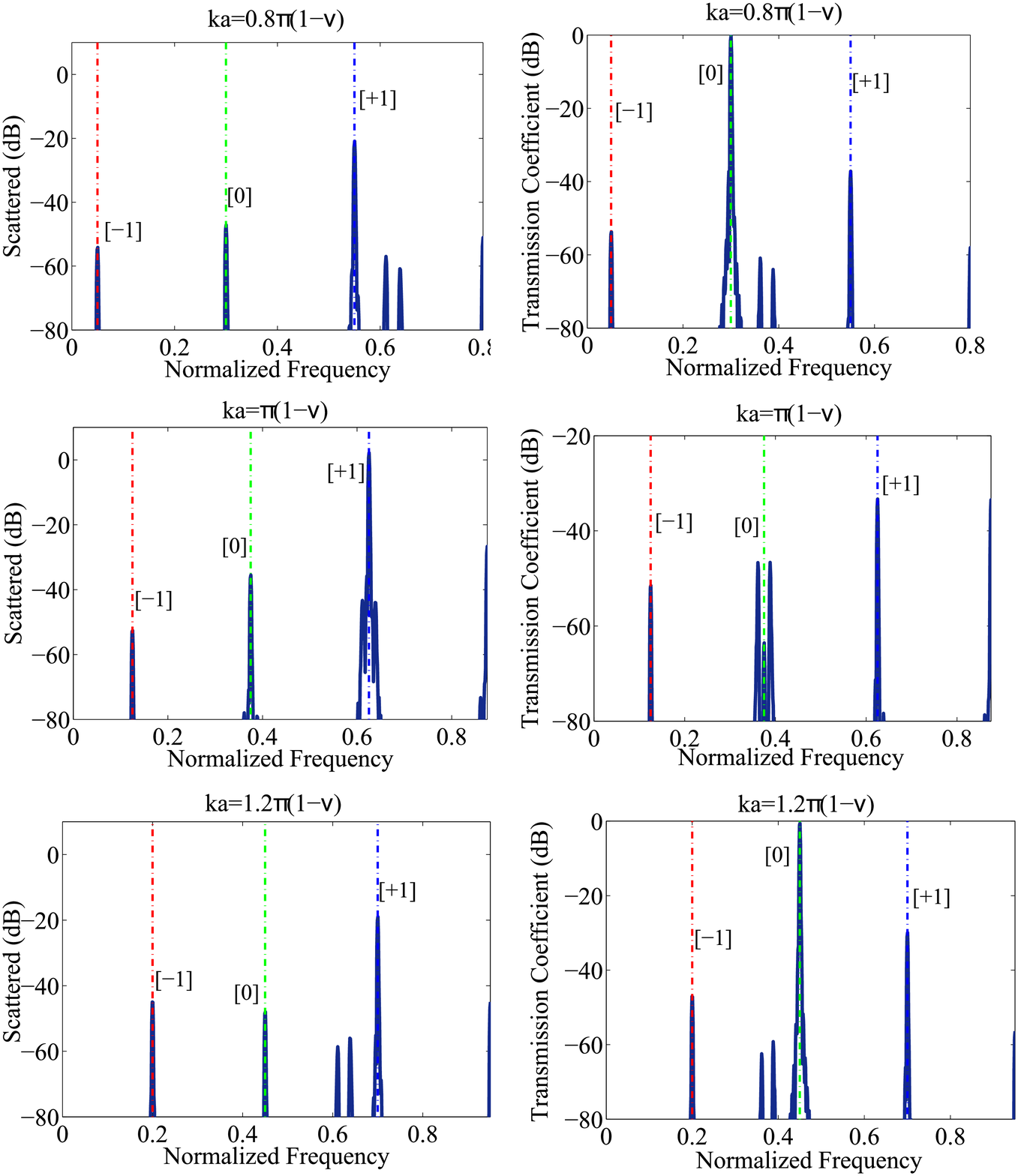}
\caption{FDTD simulated spectra for the scattered and transmitted waves for $0.8\pi(1-\nu)$, $\pi(1-\nu)$ and $1.2\pi(1-\nu)$. ([0]: incident wave at $\omega$, [-1]: $\omega-\omega_m$, [+1]: $\omega+\omega_m$).}
\label{fig:Scatter_Trans_Bwd}
\end{figure}

To demonstrate that scattering occurs only inside the band gap, Fig. \ref{fig:Scatter_Trans_Bwd} depicts three different situations, where the FDTD algorithm was used to determine the wave propagation behaviour for an input wave of $ka=0.8\pi(1-\nu)$, $ka=\pi(1-\nu)$ and $ka=1.2\pi(1-\nu)$, when $M=0.1$ and $\nu=0.1$.  The first and last frequencies are outside the bandgap. As expected, strong Brillouin-like scattering occurs when $ka=\pi(1-\nu)$ only. For the other two out-of-band frequencies ($ka=0.8\pi(1-\nu)$ and $ka=1.2\pi(1-\nu)$) the medium is transparent, demonstrated by 0~dB in the transmission spectra in Figs. \ref{fig:Scatter_Trans_Bwd}(a) and (c). 

\begin{figure}
\centering
\includegraphics[width=3.0in]{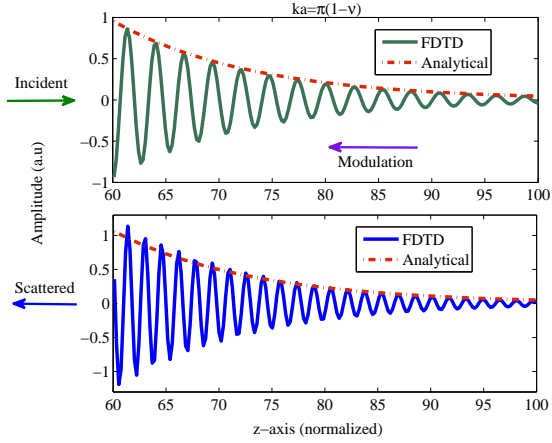}
\caption{Electric field inside the active (modulated) region for frequency component $ka=\pi(1-\nu)$ and the +1 harmonic, showing exponential conversion.}
\label{fig:growth_Bwd}
\end{figure}

According to Eqs. \ref{eq:alphaB} and \ref{eq:k_Bwd}, the field at $\omega$ is actively converted to the one at $\omega_A=\omega+\omega_m$. To demonstrate this, Fig. \ref{fig:growth_Bwd} presents the fields calculated at both frequencies. These plots are determined from the FDTD computations, after applying Fourier transform. The plots do indeed verify that the conversion is exponential; the incident fields are scattered into the $\omega+\omega_m$ frequency (blue-shifted), which bounces back to the source. Additionally, the envelopes of the two waves, determined from \ref{eq:alphaB}, \ref{eq:k_Bwd} and \ref{eq:a1_a0_Bwd} match the FDTD calculations.

\begin{figure}
\centering
\includegraphics[width=2.8in]{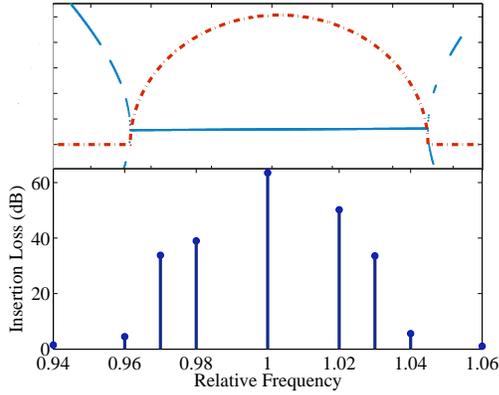}
\caption{FDTD simulated insertion loss for different frequencies in the vicinity of the band gap, superimposed on the calculated dispersion relation near the band gap.}
\label{fig:InsertionLoss_Dispersion_Bwd}
\end{figure}

As discussed earlier, the modulation results in a bandgap, where scattering is the strongest at the centre of the gap and decreases until it eventually becomes zero at the band edges. In Fig. \ref{fig:InsertionLoss_Dispersion_Bwd} we plot the FDTD calculated insertion loss for different input frequencies inside the gap, superimposed on the dispersion characteristics determined by (\ref{eq:characteristic}). As Fig. \ref{fig:InsertionLoss_Dispersion_Bwd} shows, the insertion loss is maximum at the centre of the bandgap and becomes negligibly small at the band edges.

\section{Conclusion}

A systematic analysis of the harmonics interactions present in a space-time modulated medium is carried out over a frequency range which extends from DC up to around the bandgaps in both the forward and backward directions. It is demonstrated that a passive second harmonic generation process does occur due to the singularity in the secular equation. However such behaviour is very weak and can be ignored. On the other hand, bandgaps in the forward and backward directions are created due to the active parametric interaction of an incident wave with its space-time harmonics. In this regime, the interaction can be described using a Brilliouin-like scattering process. The strength of scattering from a bandgap as well as its width were determined. To have an optimal full non-reciprocal behaviour, the modulation speed may not be below a certain threshold which is a function of the modulation index. Finally FDTD was used to verify the theoretical results and findings.

\appendix[FDTD Implementation]
\begin{figure}
\centering
\includegraphics[width=3.5in]{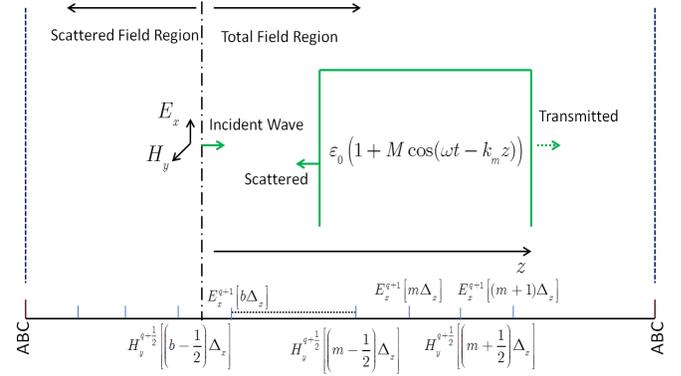}
\caption{Grid used for the FDTD solver. The space-time modulated media is excited by an incident plane wave introduced at the TF/SF interface.}
\label{fig:FDTD_geometry}
\end{figure}
 As depicted in Fig. \ref{fig:FDTD_geometry}, the medium is excited by an incident wave travelling in the $+z$ direction, which is applied at a Total Field/Scattered Field (TF/SF) interface \cite{Schneider_Online, Taflove2000}. To the right of the interface is the total field region. The grid is terminated from both sides by absorbing boundaries. The electric and magnetic fields are polarized in the $x$ and $y$ directions, respectively to guarantee that the wave propagates in the $+z$ direction. Taking into account the space time variation of the dielectric constant $\epsilon$, the FDTD discretized form of Ampere's law can be written in terms of Courant number $S_c\equiv c\Delta_t/\Delta_z$ (where $\Delta_z$ and $\Delta_t$ are the discretized spatial and temporal step, respectively) as:
 
\small
\begin{equation*}
\left(E_x\right)^{q+1}_m=\frac{\epsilon^q_m}{\epsilon^{q+1}_m}\left(E_x\right)^q_m-\frac{\eta_0S_c}{\epsilon^{q+1}_m}\left(\left(H_y \right)^{q+1/2}_{m+1/2}-\left(H_y\right)^{q+1/2}_{m-1/2} \right).
\end{equation*}
\normalsize
Here the superscript  and subscript indicate the time and spatial steps, respectively; $\eta_0$ is the impedance of free space, and $\epsilon^q_m$ is the relative permittivity at grid point $m$ at time $q\Delta_t$. The update equation of the magnetic field is obtained via the discretization of Faraday's law:

\begin{equation*}
\label{eq:updateH}
\left(H_y\right)^{q+1/2}_{m+1/2}=\left(H_y\right)^{q-1/2}_{m+1/2}-\frac{S_c}{\eta_0}\left(\left(E_x\right)^{q}_{m+1}-\left(E_x\right)^q_m\right).
\end{equation*}

The TF/SF plane is positioned between an H and E nodes. Hence the update equation for the H node at $b-1/2$ and E node at $b$ are amended by the following equations\cite{Schneider_Online}
\begin{eqnarray*}
\left(H_y\right)^{q+1/2}_{b-1/2}=\left(H_y\right)^{q+1/2}_{b-1/2}+\frac{S_c}{\eta_0}E_{ox}^\textnormal{inc}\cos(\omega q\Delta_t)\\
\left(E_x\right)^{q+1}_b=\left(E_x\right)^{q+1}_b+\frac{S_c}{\epsilon^{q+1}_b}E_{ox}^\textnormal{inc}\cos(\omega(q+1/2)\Delta_t+k\Delta x_2),
\end{eqnarray*}
where $E_{ox}^\textnormal{inc}$ is the electric field of the incident wave, $\omega$ and $k$ are its frequency and wave number, respectively. In the above eqns., the incident electric field is referenced to the position $b\Delta_z$ (i.e, $E=E_{ox}^\textnormal{inc}\cos(\omega t-k[z-b])$).
% if have a single appendix:
%\appendix[Proof of the Zonklar Equations]
% or
%\appendix  % for no appendix heading
% do not use \section anymore after \appendix, only \section*
% is possibly needed

% use appendices with more than one appendix
% then use \section to start each appendix
% you must declare a \section before using any
% \subsection or using \label (\appendices by itself
% starts a section numbered zero.)
%

%\appendices
%\section{Proof of the First Zonklar Equation}
%Appendix one text goes here.

% you can choose not to have a title for an appendix
% if you want by leaving the argument blank
%\section{}
%Appendix two text goes here.

% use section* for acknowledgement
%\section*{Acknowledgment}

%The authors would like to thank...

% Can use something like this to put references on a page
% by themselves when using endfloat and the captionsoff option.
%\ifCLASSOPTIONcaptionsoff
  %\newpage
%\fi

\bibliographystyle{IEEEtran}

\bibliography{IEEEabrv,Nonreci}

% Generated by IEEEtran.bst, version: 1.14 (2015/08/26)
\begin{thebibliography}{10}
\providecommand{\url}[1]{#1}
\csname url@samestyle\endcsname
\providecommand{\newblock}{\relax}
\providecommand{\bibinfo}[2]{#2}
\providecommand{\BIBentrySTDinterwordspacing}{\spaceskip=0pt\relax}
\providecommand{\BIBentryALTinterwordstretchfactor}{4}
\providecommand{\BIBentryALTinterwordspacing}{\spaceskip=\fontdimen2\font plus
\BIBentryALTinterwordstretchfactor\fontdimen3\font minus
  \fontdimen4\font\relax}
\providecommand{\BIBforeignlanguage}[2]{{%
\expandafter\ifx\csname l@#1\endcsname\relax
\typeout{** WARNING: IEEEtran.bst: No hyphenation pattern has been}%
\typeout{** loaded for the language `#1'. Using the pattern for}%
\typeout{** the default language instead.}%
\else
\language=\csname l@#1\endcsname
\fi
#2}}
\providecommand{\BIBdecl}{\relax}
\BIBdecl

\bibitem{Tien1958}
P.~Tien, ``Parametric amplification and frequency mixing in propagating
  circuits,'' \emph{Journal of Applied Physics}, vol.~29, no.~9, pp.
  1347--1357, 1958.

\bibitem{Cullen1958}
A.~Cullen, ``A travelling-wave parametric amplifier,'' \emph{Nature}, vol. 181,
  no. 4605, pp. 332--332, 1958.

\bibitem{Simon1960}
J.-C. Simon, ``Action of a progressive disturbance on a guided electromagnetic
  wave,'' \emph{IRE Transactions on Microwave Theory and Techniques}, vol.~8,
  no.~1, pp. 18--29, 1960.

\bibitem{Oliner1961}
A.~A. Oliner and A.~Hessel, ``Wave propagation in a medium with a progressive
  sinusoidal disturbance,'' \emph{IRE Transactions on Microwave Theory and
  Techniques}, vol.~9, no.~4, pp. 337--343, July 1961.

\bibitem{Cassedy1963}
E.~S. Cassedy and A.~A. Oliner, ``Dispersion relations in time-space periodic
  media: Part i; stable interactions,'' \emph{Proceedings of the IEEE},
  vol.~51, no.~10, pp. 1342--1359, Oct 1963.

\bibitem{Cassedy1967}
E.~S. Cassedy, ``Dispersion relations in time-space periodic media part ii;
  unstable interactions,'' \emph{Proceedings of the IEEE}, vol.~55, no.~7, pp.
  1154--1168, July 1967.

\bibitem{Boyd}
R.~W. Boyd, \emph{Nonlinear optics}, 3rd~ed.\hskip 1em plus 0.5em minus
  0.4em\relax Boston: Boston : Academic Press, 2008, includes bibliographical
  references and index.

\bibitem{Fabelinskii}
I.~L. Fabelinskii, \emph{Molecular scattering of light}.\hskip 1em plus 0.5em
  minus 0.4em\relax Springer Science \& Business Media, 2012.

\bibitem{Slater1958}
J.~C. Slater, ``Interaction of waves in crystals,'' \emph{Reviews of Modern
  Physics}, vol.~30, no.~1, p. 197, 1958.

\bibitem{Sameh_JAP_NLD}
S.~Y. Elnaggar and G.~N. Milford, ``Description and stability analysis of
  nonlinear transmission line type metamaterials using nonlinear dynamics
  theory,'' \emph{Journal of Applied Physics}, vol. 121, no.~12, p. 124902,
  2017.

\bibitem{Oliner1959}
A.~Oliner and A.~Hessel, ``Guided waves on sinusoidally-modulated reactance
  surfaces,'' \emph{IRE Transactions on Antennas and Propagation}, vol.~7,
  no.~5, pp. 201--208, 1959.

\bibitem{Sameh_TAP_TWM}
S.~Elnaggar and G.~Milford, ``Three wave mixing as the limit of nonlinear
  dynamics theory for nonlinear transmission line type metamaterials,''
  \emph{IEEE Transactions on Antennas and Propagation, Under review}.

\bibitem{Fan2009}
Z.~Yu and S.~Fan, ``Complete optical isolation created by indirect interband
  photonic transitions,'' \emph{Nat Photon}, vol.~3, no.~2, pp. 91--94, 2009,
  10.1038/nphoton.2008.273.

\bibitem{Lira2012}
H.~Lira, Z.~Yu, S.~Fan, and M.~Lipson, ``Electrically driven nonreciprocity
  induced by interband photonic transition on a silicon chip,'' \emph{Phys.
  Rev. Lett.}, vol. 109, p. 033901, Jul 2012.

\bibitem{AluCaloz}
D.~L. Sounas, C.~Caloz, and A.~Alù, ``Giant non-reciprocity at the
  subwavelength scale using angular momentum-biased metamaterials,''
  \emph{Nature Communications}, vol.~4, p. 2407, 2013.

\bibitem{Winn1999}
J.~N. Winn, S.~Fan, J.~D. Joannopoulos, and E.~P. Ippen, ``Interband
  transitions in photonic crystals,'' \emph{Phys. Rev. B}, vol.~59, pp.
  1551--1554, Jan 1999.

\bibitem{Hadad2016}
Y.~Hadad, J.~C. Soric, and A.~Alu, ``Breaking temporal symmetries for emission
  and absorption,'' \emph{Proceedings of the National Academy of Sciences},
  vol. 113, no.~13, pp. 3471--3475, 2016.

\bibitem{Taravati2017mixer}
S.~Taravati and C.~Caloz, ``Mixer-duplexer-antenna leaky-wave system based on
  periodic space-time modulation,'' \emph{IEEE Transactions on Antennas and
  Propagation}, vol.~65, no.~2, pp. 442--452, 2017.

\bibitem{Qin2014}
S.~Qin, Q.~Xu, and Y.~E. Wang, ``Nonreciprocal components with distributedly
  modulated capacitors,'' \emph{IEEE Transactions on Microwave Theory and
  Techniques}, vol.~62, no.~10, pp. 2260--2272, 2014.

\bibitem{Alu2016magnetless}
N.~A. Estep, D.~L. Sounas, and A.~Al{\`u}, ``Magnetless microwave circulators
  based on spatiotemporally modulated rings of coupled resonators,'' \emph{IEEE
  Transactions on Microwave Theory and Techniques}, vol.~64, no.~2, pp.
  502--518, 2016.

\bibitem{Taravati2017oblique}
S.~Taravati, N.~Chamanara, and C.~Caloz, ``Nonreciprocal electromagnetic
  scattering from a periodically space-time modulated slab and application to a
  quasi-sonic isolator,'' \emph{arXiv:1705.06311}, 2017.

\bibitem{Caloz2016_STmetasurfaces}
C.~Caloz, K.~Achouri, Y.~Vahabzadeh, and N.~Chamanara, ``Spacetime
  metasurfaces,'' in \emph{2016 Photonics North (PN)}, May 2016, pp. 1--2.

\bibitem{Trainiti2016}
G.~Trainiti and M.~Ruzzene, ``Non-reciprocal elastic wave propagation in
  spatiotemporal periodic structures,'' \emph{New Journal of Physics}, vol.~18,
  no.~8, p. 083047, 2016.

\bibitem{Croenne2017brillouin}
C.~Cro{\"e}nne, J.~Vasseur, O.~Bou~Matar, M.-F. Ponge, P.~Deymier, A.-C.
  Hladky-Hennion, and B.~Dubus, ``Brillouin scattering-like effect and
  non-reciprocal propagation of elastic waves due to spatio-temporal modulation
  of electrical boundary conditions in piezoelectric media,'' \emph{Applied
  Physics Letters}, vol. 110, no.~6, p. 061901, 2017.

\bibitem{Louisell1960}
W.~H. Louisell.\hskip 1em plus 0.5em minus 0.4em\relax New York, Wiley, 1960,
  (William Henry).

\bibitem{Pierce1954}
J.~R. Pierce, ``Coupling of modes of propagation,'' \emph{Journal of Applied
  Physics}, vol.~25, no.~2, 1954.

\bibitem{Schneider_Online}
J.~B. Schneider, ``Understanding the finite-difference time-domain method,''
  \emph{School of electrical engineering and computer science Washington State
  University.--URL: http://www. Eecs. Wsu. Edu/\~{} schneidj/ufdtd/(request
  data: 29.11. 2012)}, 2010.

\bibitem{Taflove2000}
A.~Taflove and S.~C. Hagness, \emph{Computational electrodynamics}.\hskip 1em
  plus 0.5em minus 0.4em\relax Artech house publishers, 2000.

\end{thebibliography}

\end{document}